\documentstyle[prl,aps,epsf,rotate,multicol]{revtex}

\begin{document}
\draft

\title{Crossover from Scale-Free to Spatial Networks}

\author{Marc Barth\'elemy}

\address{CEA, Service de Physique de la Mati\`ere Condens\'ee\\
BP12, 91680 Bruy\`eres-le-Ch\^atel, France
}

\date{\today}
\maketitle
\begin{abstract}
In many networks such as transportation or communication networks,
distance is certainly a relevant parameter. In addition, real-world
examples suggest that when long-range links are existing, they usually
connect to hubs-the well connected nodes. We analyze a simple model
which combine both these ingredients---preferential attachment and
distance selection characterized by a typical finite `interaction
range'. We study the crossover from the scale-free to the `spatial'
network as the interaction range decreases and we propose scaling
forms for different quantities describing the network. In particular,
when the distance effect is important (i) the connectivity
distribution has a cut-off depending on the node density, (ii) the
clustering coefficient is very high, and (iii) we observe a positive
maximum in the degree correlation (assortativity) which numerical
value is in agreement with empirical measurements. Finally, we show
that if the number of nodes is fixed, the optimal network which
minimizes both the total length and the diameter lies in between the
scale-free and spatial networks. This phenomenon could play an
important role in the formation of networks and could be an
explanation for the high clustering and the positive assortativity
which are non trivial features observed in many real-world examples.
\end{abstract}


\begin{multicols}{2}


Even if some networks are defined without any reference to an
embedding space, it is not the case for most real-world networks. Most
people have their friends and relatives in their neighborhood,
transportation networks depend obviously on distance, many
communication networks devices have short radio
range\cite{Helmy02,Nemeth02,Gorman03}, and the spread of contagious
diseases is not uniform across territories. A particularly important
example of such a spatial network is the Internet which is defined as
a set or routers linked by physical cables with different lengths and
latency times \cite{Lakhina02}. From these examples, it appears
important to define a realistic model in which nodes and links are
embedded in a physical space which induces a distance between nodes,
and we will designate these networks as `spatial'. More generally, the
distance could be another parameter such as a social distance measured
by salary, socio-professional category differences, or any quantity
which measures the cost associated with the link formation. If the
cost of a long-range link is high, most of the connections starting
from a given node will link to the nearest neighbors. When a
long-range link is existing, it will usually connect to a
well-connected node---that is, a hub. This is for instance the case
for airlines: Short connections go to small airports while long
connections point preferably to big airport ({\it ie.} well connected
nodes). This propensity to link to an already well connected node has
been coined preferential attachment\cite{Simon55,Barabasi99}: The
probability to link to a node is proportional to the connectivity of
it. It is widely accepted that preferential attachment is the probable
explanation for the power-law distribution seen in many
networks\cite{Barabasi99}. However, even if this process generates
reasonably realistic networks, it misses the important element of the
cost of links. We will study a simple model which incorporates such a
cost with a typical scale $r_c$.

Several models including distance were previously proposed
\cite{Yook02,Rozen02,Warren02,Dall02,Sen02,Jost02,Manna02,Xulvi02} but
the case of preferential attachment with a finite scale $r_c$ was not
considered before and more generally, the study of the interplay
between preferential attachment and distance effects is still
lacking. In this Letter, we study this interplay and we demonstrate
the existence of a crossover from the scale-free network to the
spatial network when the interaction range decreases. In particular,
we propose scaling forms for the different quantities which
characterize the network.


The $N$ nodes of the network are supposed to be in a $d$-dimensional
space of linear size $L$ and we will assume that they are distributed
randomly in space with uniform density $\rho$. One could use other
distributions: For instance in cities the density decreases
exponentially from the center\cite{Clark51}. The case of randomly
distributed points is interesting since on average it preserves
natural symmetries such as translational and rotational invariance in
contrast with lattices. For the sake of simplicity, we will choose for
our numerical simulations the two-dimensional plane and the Euclidean
distance. Once the nodes are distributed in this space, we have to
construct the links and we use the following algorithm: (1) Select at
random a subset of $n_0$ initial active nodes. (2) Take an inactive
node $i$ at random an connect it with an active node $j$ with
probability (up to a normalization factor)
\begin{eqnarray}
p_{i\to j}\propto\frac{Z(k_j)}{\Delta(d_{ij})}
\end{eqnarray}
where $k_j$ is the connectivity of node $j$, $d_{ij}$ is the distance
between nodes $i$ and $j$, and $Z$ and $\Delta$ are given
functions. Finally (3), make the node $i$ active and go back to $(2)$
until all nodes are active. For each node, we repeat $m$ times the
steps $(2-3)$ so that the average connectivity will be $\langle
k\rangle=2m$ (numerically we choose $m=3$). There are essentially
three different interesting cases: (i) {\it Preferential attachment}:
When $Z(k)=k+1$, and $\Delta={\rm const.}$, we recover the usual
preferential attachment problem\cite{Barabasi99}: The connectivity
distribution is a power law with exponent $\gamma=3$, the shortest
path $\ell$ is growing with $N$ as $\ell\sim\log N$, the
assortativity is decreasing as $\log^2N/N$\cite{Newman02a}, and the
clustering coefficient is decreasing as $1/N$\cite{Klemm02}. (ii) {\it
Distance selection}: In this case $Z={\rm const.}$: There is a
distance effect only\cite{Nemeth02,Dall02,Sen02,Jost02}. Jost and
Joy\cite{Jost02} studied different functions $\Delta(d)$, Dall and
Christensen\cite{Dall02} studied graphs in which each vertex is
randomly located and connected with its adjacent points, while
in\cite{Sen02}, the authors study a (small-world) network constructed
from re-wiring links with a probability that decreases as a power-law
with distance. (iii) {\it Preferential attachment and distance
selection}: It is the case where $Z(k)=k+1$ and $\Delta$ is a
increasing function of the distance. In most cases---such as
transportation networks or social interactions---the range of
interaction is limited, which is explained by the fact that there is a
cost associated to long range links.  Up to our knowledge, studies
done so far in this case were concerned with the case where $\Delta$
decreases as a power-law \cite{Manna02,Xulvi02,Yook02}. The main
result is then the existence of different regimes according to the
value of the exponent describing the spatial decay of $\Delta$. In
contrast, in this Letter we will study the finite-range case for which
the function $\Delta$ is negligible above a finite scale $r_c$, a
prototype being the exponential function
\begin{equation}
\Delta(d)=e^{d/r_c}
\label{exp}
\end{equation}
Even if there is some controversy on the spatial decay of the linking
probability for Internet cables\cite{Lakhina02,Yook02}, it seems that
the range is relatively short and that when a new server (or router)
adds to the network, it will connect preferably to the nearest
node(s). One of the most important model of Internet topology relying
on this argument and using Equ.~(\ref{exp}) is the Waxman topology
generator\cite{Waxman88} and the model considered here thus appears
has the natural generalization of the Waxman case.


When the interaction range is of the order of the system size (or
larger), we expect the distance to be irrelevant and the obtained
network will be scale-free. In contrast, when the interaction range is
small compared to the system size, we expect new properties and we
would like to understand the crossover between these two regimes as
well as the scaling for the different quantities describing the
network.

\begin{figure}
\narrowtext
\centerline{
\epsfysize=0.4\columnwidth{\epsfbox{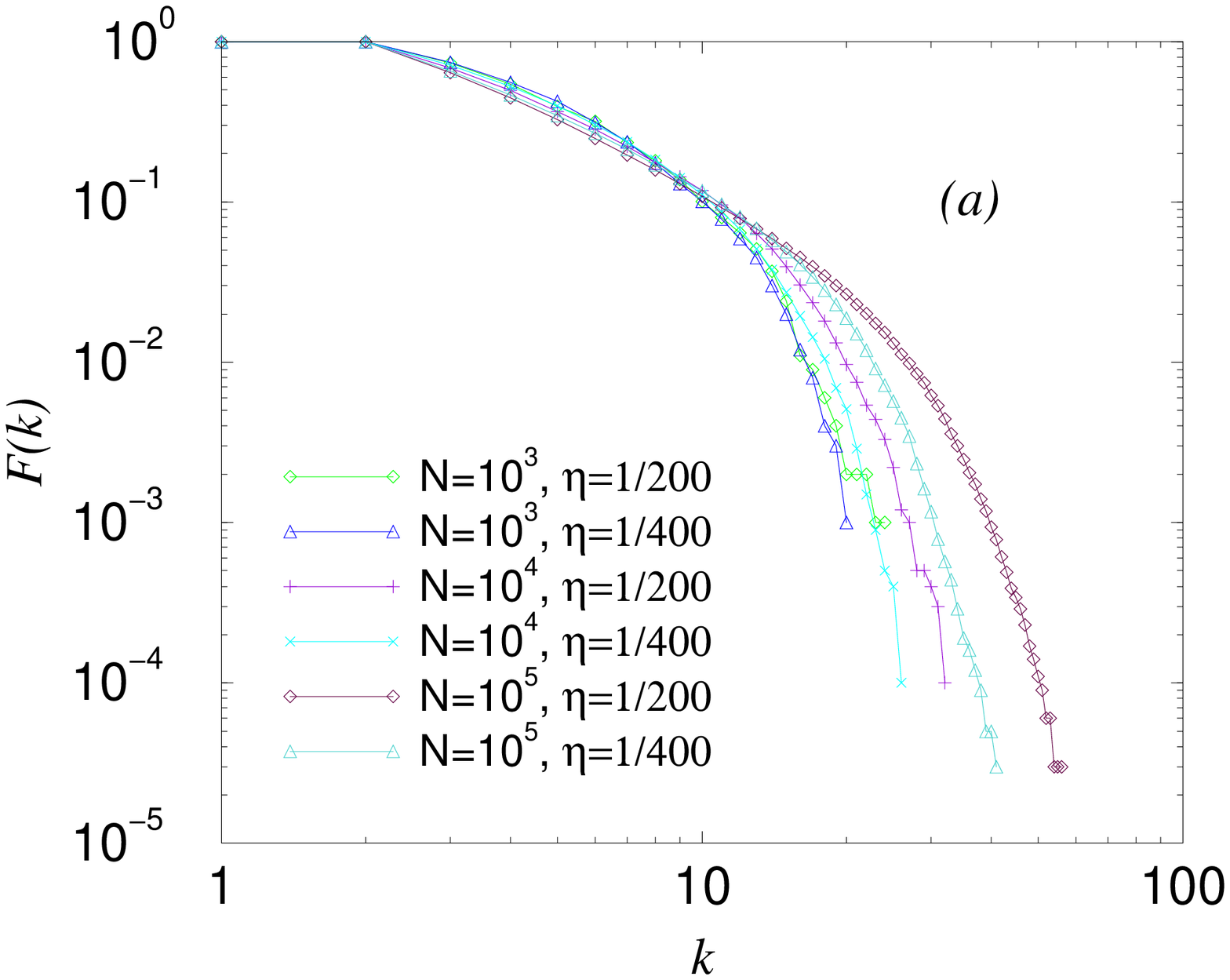}}
\hspace*{0.01cm}
\epsfysize=0.4\columnwidth{\epsfbox{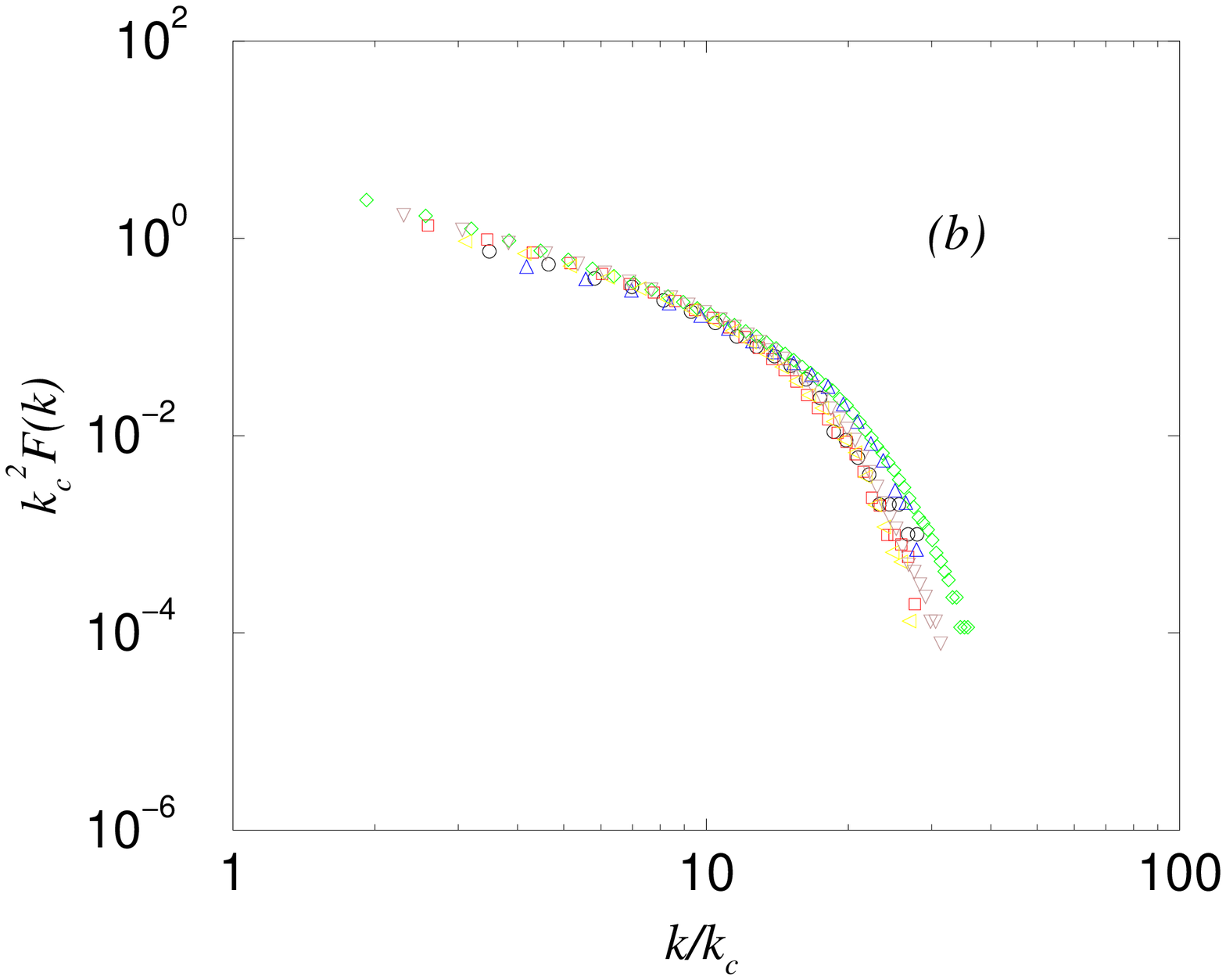}}
}
\caption{ (a) Cumulative distribution function for different values of
$N$ going from $N=10^3$ to $10^5$ and for $\eta$ going from $1/400$ to
$1/200$. (b) Data collapse for the same data of figure (b) with
$k_c\sim n^\beta$ with $\beta\simeq 0.13$.}
\label{fk.fig}
\end{figure}

{\it Probability distribution}. We did simulations for this model and
as expected, when $\eta\equiv r_c/L$ is larger than one, the distance
selection is in-operant and we recover the usual scale-free network
obtained by preferential attachment. In particular, the probability
distribution is a power law with exponent $\gamma=3$ independently of
the actual value of $\eta$ (not shown). In the opposite
situation $\eta\ll 1$, the distance is relevant and we expect a
cut-off in the (one minus the) cumulative connectivity distribution
\begin{equation}
F(k)\sim k^{-\gamma+1}f(\frac{k}{k_c})
\label{fk}
\end{equation}
where $k_c$ is the cut-off at large connectivity: $f(x\gg 1)\simeq 0$
(see figure \ref{fk.fig}a). In this problem, we have two
dimension-less parameters: The number of nodes $N$ and $\eta$. Our
guess---a posteriori verified---is that the control parameter is the
average number of points present in a sphere of radius $r_c$ (and of
volume $V_d(r_c)$) as given by $n=\rho V_d(r_c)$. We thus propose the
following scaling ansatz for the cut-off valid only for $\eta\ll 1$
\begin{equation}
k_c\sim n^{\beta}
\label{kc}
\end{equation}
In order to test these assumptions, we first plot the cumulative
distribution for different values of $N$, $r_c$ and $L$ for $d=2$ (see
Fig.~\ref{fk.fig}b). We then use our scaling ans\"atze
(\ref{fk}),(\ref{kc}) and we indeed obtain a good data collapse
(Fig.~\ref{fk.fig}c) with $\beta\simeq 0.13$. According to these
results, the distance effect limits the choice of available
connections thereby limiting the connectivity distribution for large
values. More generally, the distance effect induces correlations which
are known to affect the connectivity distribution\cite{Berg02}.

\begin{figure}
\narrowtext
\centerline{
\epsfysize=0.4\columnwidth{\epsfbox{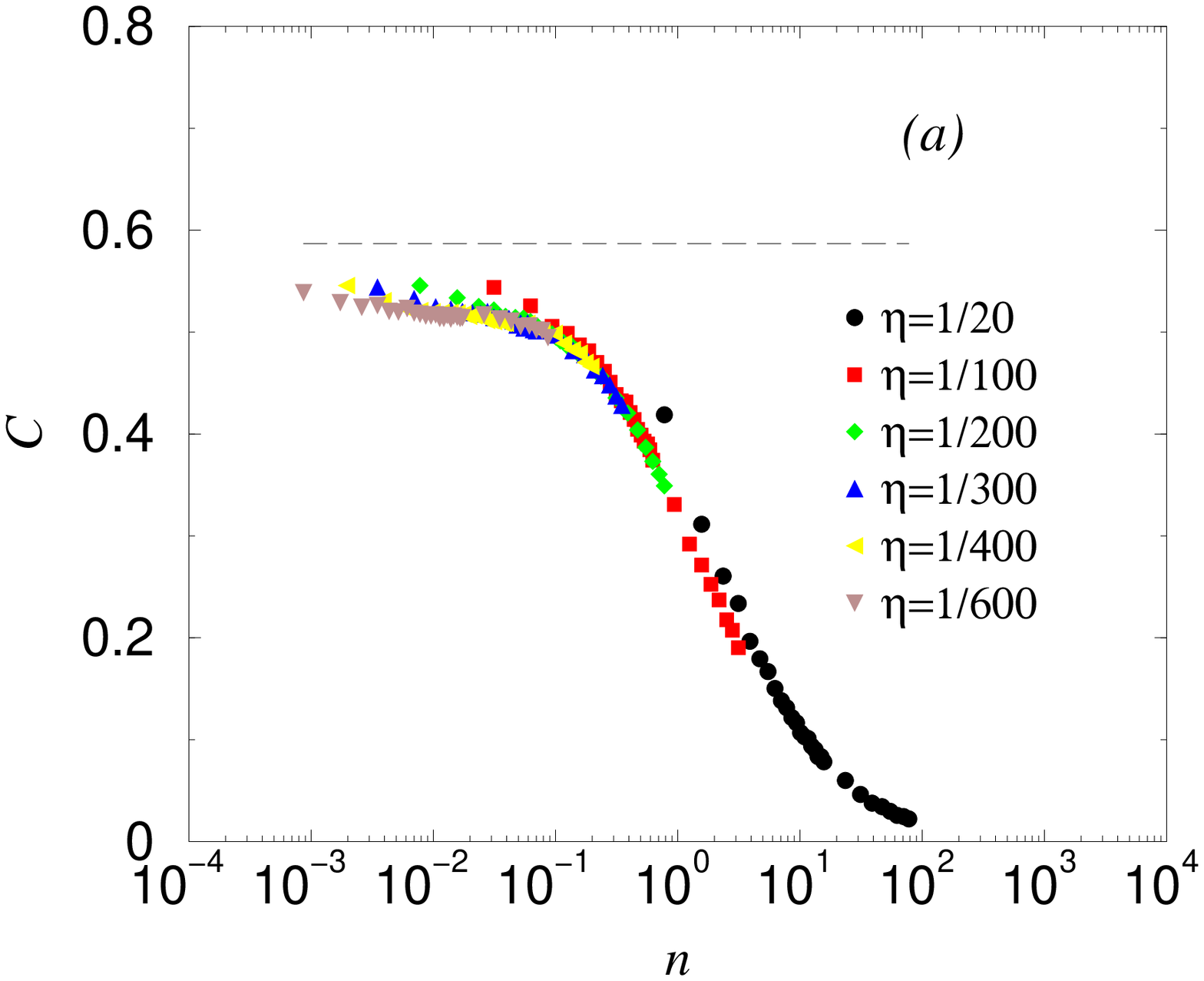}}
\epsfysize=0.4\columnwidth{\epsfbox{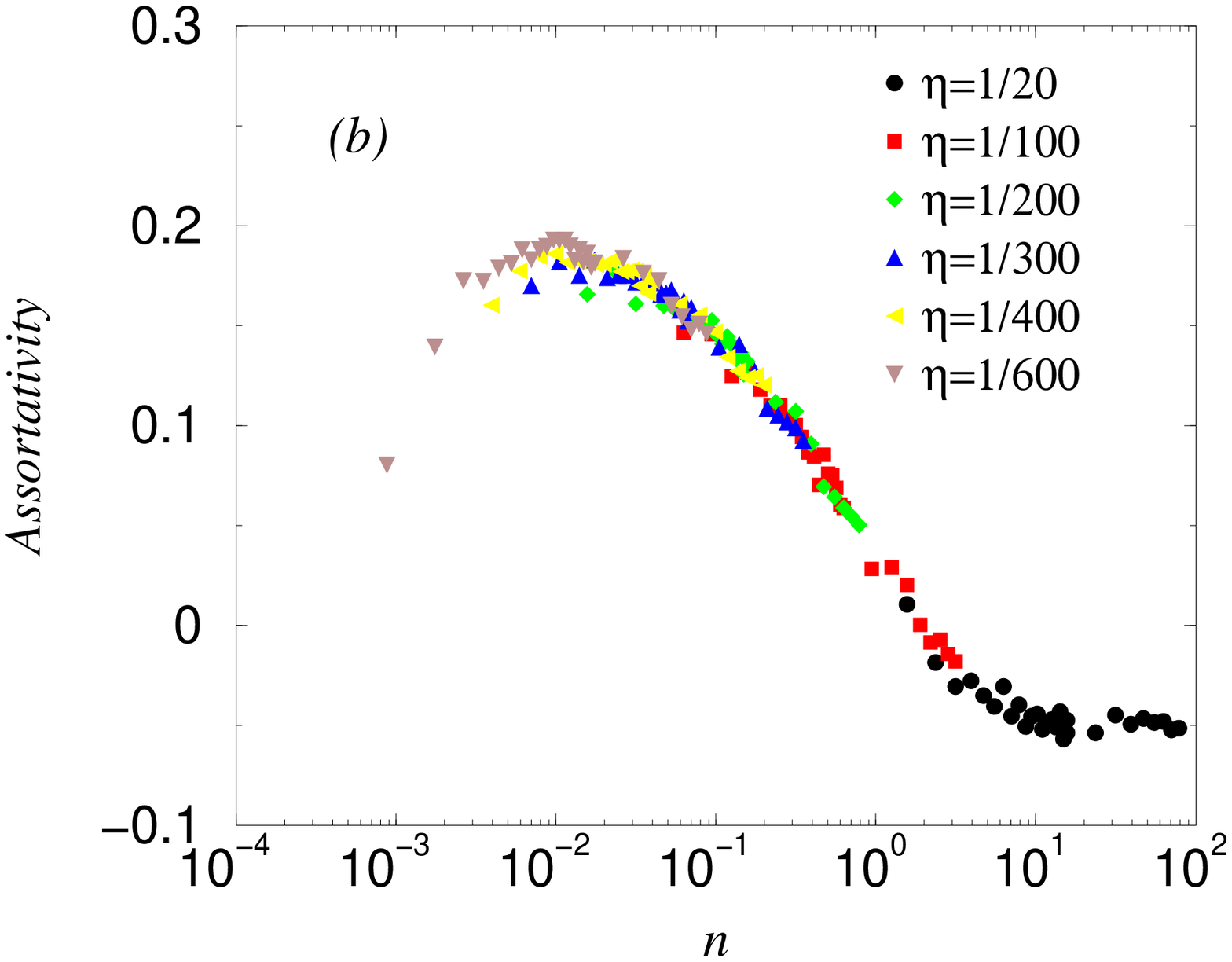}}
}
\vspace*{0.1cm}
\caption{ Clustering coefficient versus versus the mean number
$n=\rho\pi r_c^2$ of points in the disk of radius $r_c$ (plotted in
Log-Lin). The dashed line corresponds to the theoretical value $C_0$
computed when a vertex connects to its adjacent neighbors without
preferential attachment $[9]$. (b) Assortativity versus $n$ (in
Log-Lin). This plot shows clearly a positive maximum for $n\simeq
0.1$. Around this maximum, the network will be resilient even to a
targeted attack against hubs.
}
\label{cv-assor}
\end{figure}

{\it Clustering coefficient}. The clustering coefficient $C$ is
defined as the average over all nodes of the percentage of existing
links between neighbors\cite{Watts98}. For the scale-free network $C$
is small and decreases as $1/N$. In contrast, when the distance effect
is important we expect a higher cluster coefficient.  Indeed, if two
given nodes $i$ and $j$ are connected it means that the distance
$d_{ij}$ is less or of the order of $r_c$. In the process of adding
links, if a new node $k$ links to $i$, it means also that
$d_{ki}<r_c$. This implies that $j$ and $k$ belongs to the disk of
center $i$ and of radius $r_c$. The probability that $k$ and $j$ will
link will depend on the distance $d_{jk}$. When we can neglect the
preferential attachment and if we suppose that a given node is
connected to all its neighbors, the probability that $k$ is also
linked to node $j$ is given in terms of the area of the intersection
of the the two spheres of radius $r_c$ centered on $i$ and $j$
respectively. This is a simple calculation done in\cite{Dall02} and
predicts that for $d=2$ the clustering coefficient is
$C_0=1-3\sqrt{3}/4\pi\simeq 0.59$. We expect to recover this limit for
$\eta\to 0$ and for an average connectivity $\langle k\rangle=6$ which
is a well-known result in random geometry\cite{Note_Poisson}. If
$\eta$ is not too small, the preferential attachment is important and
induces some dependence of the clustering coefficient on $N$. In
addition, we expect that $C$ will be lower than $C_0$ since in our
model the links don't connect necessarily to the nearest neighbors. We
first compute $C$ versus the number of nodes $N$ for different values
of $\eta$ (not shown) and the same data versus $n$
(Fig.~\ref{cv-assor}a) fall nicely on the same curve which is a
decreasing function. In order to understand this variation, we
consider one given node $i_0$. When $n$ is above $1$ and increasing,
the number of neighbors of the node $i_0$ will increase and the
probability that two of them will be linked is thus decreasing which
explains the monotonic decay of $C(n)$.


{\it Assortativity}. We also compute the assortativity $A$ of the network
which measures the correlation between the degree of the nodes. In
this case, one way to measure this correlation is the Pearson
correlation coefficient of the degrees at either ends of an
edge\cite{Callaway01}. This quantity varies from $-1$ (disassortative
network) to $1$ (perfectly assortative network) and is decreasing
towards zero as $\log^2N/N$ for the scale-free network. In the case
$\eta\ll 1$, we compute the assortativity coefficient versus $N$ and
we plot the results versus $n$ (Fig.~\ref{cv-assor}b). The data fall
onto a single curve and exhibit a maximum for $n\simeq 0.1$ where the
network is also very clustered. In addition, this maximum is positive
which indicates that the hubs are connected and not dispersed in the
network. For very small $n$, the connectivity does not fluctuate and
$A\simeq 0$ while for large $n$ preferential attachment dominates and
$A$ is also small. This could be a possible explanation for the
maximum observed in the intermediate regime where distance selection
and preferential attachment coexist. It is interesting to note that
the value of the maximum ($A_{\rm max}\simeq 0.2$) is in agreement
with the empirical measurements for different networks\cite{Newman02b}
which give $A\in[0.20, 0.36]$. The cost of links thus induces
positive correlations and could provide a simple explanation to the
positive assortativity observed in social networks (actors,
collaboration, or co-authorship). This fact is also very important
from the point of view of the resilience of the network since it is
very sensitive to the degree correlation\cite{Newman02b}. In
particular, for the scale-free network, deleting hubs when $A>0$ is
not as efficient when $A\simeq 0$ or $A<0$\cite{Vasquez02}. This means
that for this type of networks, one has to adapt the best attack
strategy to the value of the density.


{\it Diameter}. An important characterization of a network is its
diameter $\ell$\cite{Bollobas}: It is the shortest path between two
nodes averaged over all pairs of nodes and counts the number of hops
between two points. 

\begin{figure}
\narrowtext
\centerline{
\epsfysize=0.4\columnwidth{\epsfbox{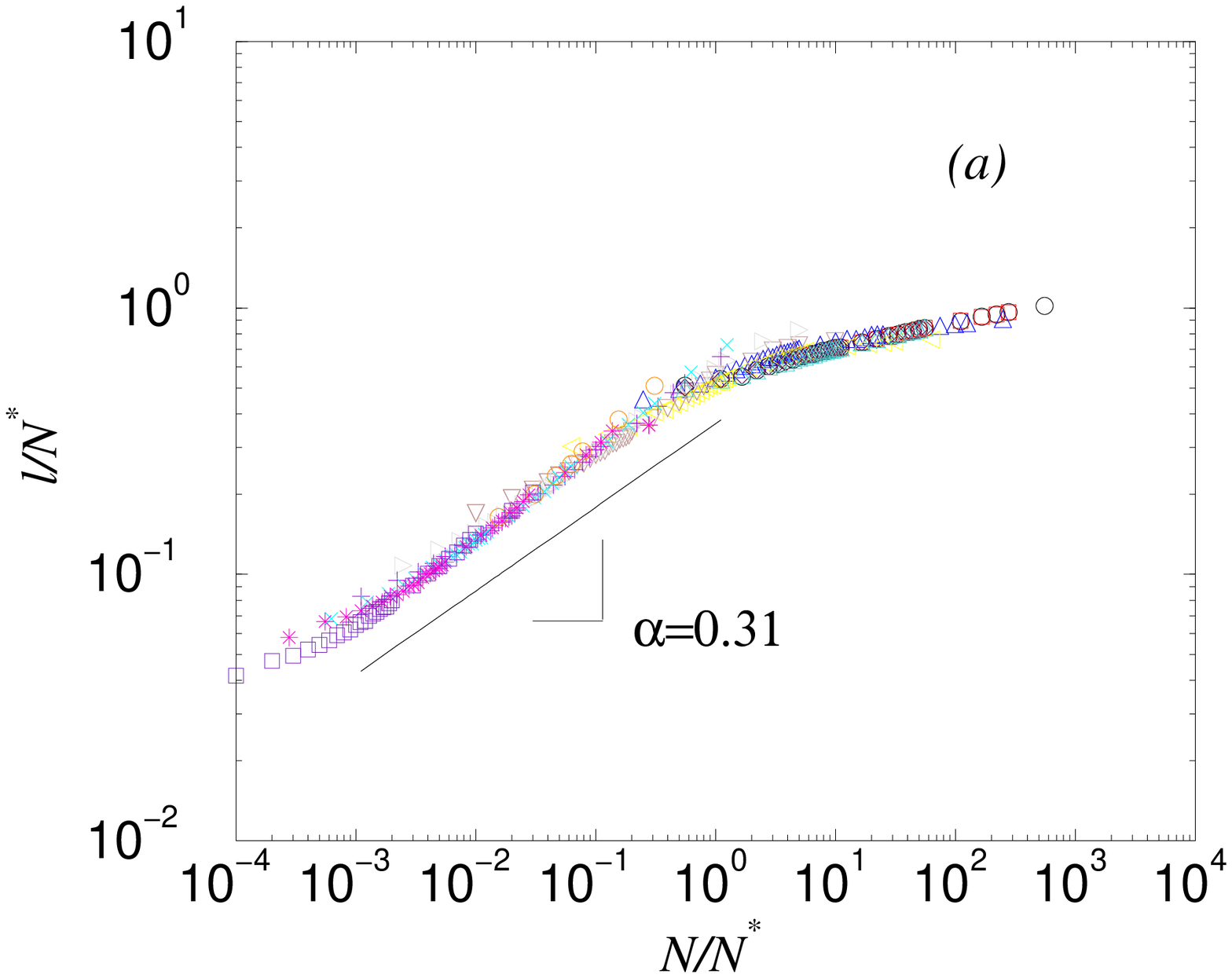}}
\hspace*{0.01cm} 
\epsfysize=0.4\columnwidth{\epsfbox{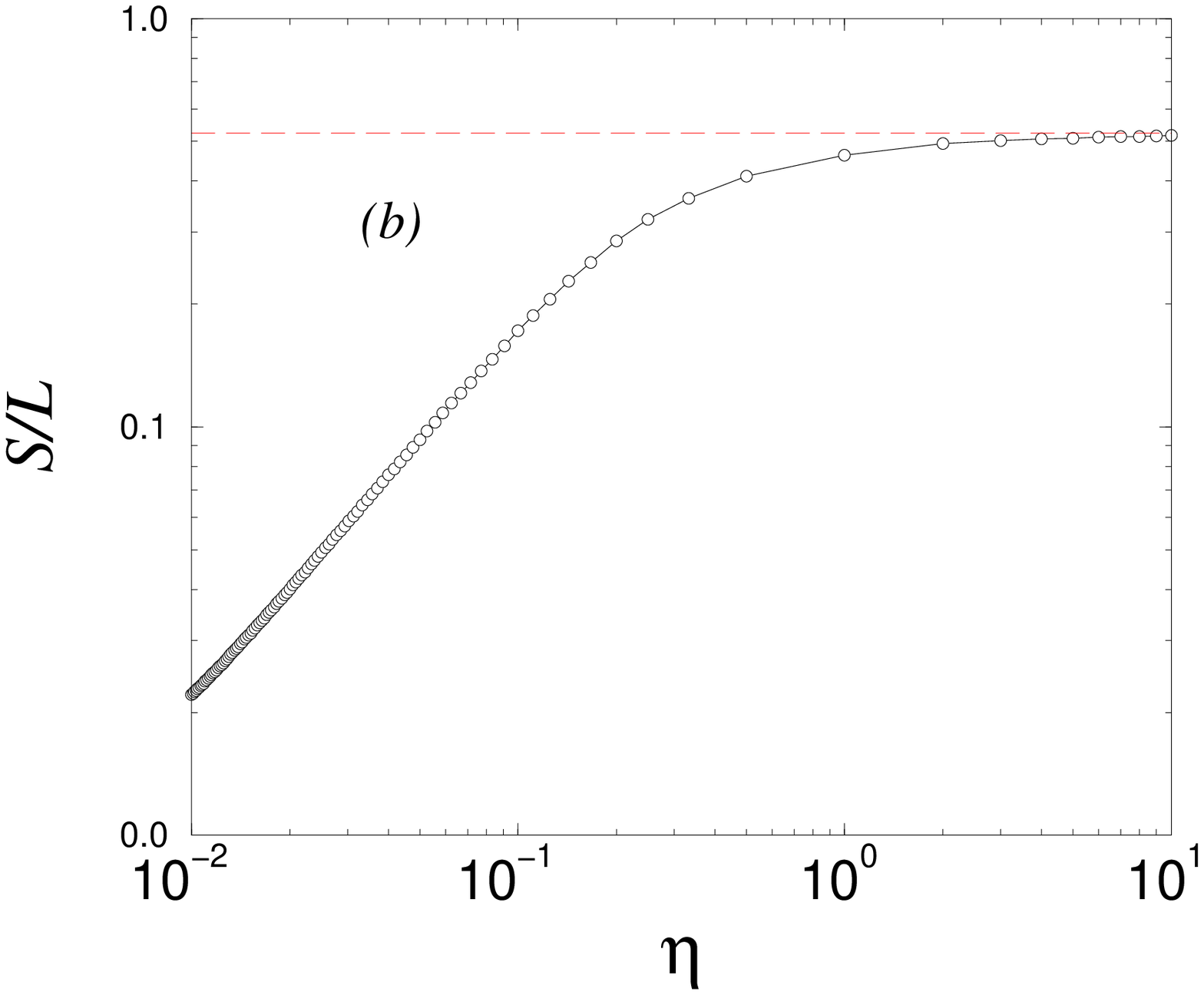}}
}
\vspace*{0.1cm}
\caption{ (a) Data collapse (in Log-Log) for $\ell(N,\eta)$ using our
scaling ansatz Equ.~(\ref{ansatz}) together with
Equs.~(\ref{eta_big}), (\ref{eta_small}). The data collapse is
obtained with $14$ curves for $\eta$ going from $1/500$ to $100$ and
for $N$ up to $10^5$. The first part of the scaling function exhibits
a power law behavior with exponent $\alpha\simeq 0.32$, followed by a
logarithmic behavior for $N/N^*\gg 1$. (b) Total length of the network
per node (and per distance) $S/L$ versus $\eta$ going from $1/100$ to
$10$. This cost is first linearly increasing with $\eta$ and as the
preferential attachment becomes more important it converges to $\pi/6$
(long-dashed line).  }
\label{ell-cost}
\end{figure}

We propose the following scaling ansatz which describe the crossover
from a spatial to a scale-free network
\begin{equation}
\ell(N,\eta)=[N^*(\eta)]^\alpha\Phi\left[\frac{N}{N^*(\eta)}\right]
\label{ansatz}
\end{equation}
with $\Phi(x\ll 1)\sim x^\alpha$ and $\Phi(x\gg 1)\sim\log x$. The
typical size $N^*$ is depending on $\eta$ and its behavior is a priori
complex. However, we can find its behavior in two extreme cases. For
$\eta\gg 1$, space is irrelevant and
\begin{equation}
N^*(\eta\gg 1)\sim N_0
\label{eta_big}
\end{equation}
where $N_0$ is a finite constant. When $\eta\ll 1$, the existence of
long-range links will determine the behavior of $\ell$. If we denote
by $a=1/\rho^{1/d}$ the typical inter-node distance, we have to
distinguish two regimes: If $r_c\ll a$ then long-range links cannot
exist and therefore $N\ll N^*$. If $r_c\gg a$, long-range links can
exist and we are in a small-world regime $N\gg N^*$. This argument
implies that $N^*$ is such that $r_c\sim a$ which in turn implies
\begin{equation}
N^*(\eta\ll 1)\sim \frac{1}{\eta^d}
\label{eta_small}
\end{equation}
In Fig.~\ref{ell-cost}a, we use the ansatz Equ.~(\ref{ansatz}) together with
the results Equ.~(\ref{eta_big}), (\ref{eta_small}). The data are
collapsing onto a single curve showing the validity of our scaling
ansatz. This data collapse is obtained for $\alpha\simeq 0.31$ and
$N_0\simeq 180$ (for $d=2$). The scale-free network is a `small'
world: the diameter is growing with the number of points as
$\ell\sim\log N$. In the opposite case of the spatial network with a
small interaction range, the network is much larger: To go from a
point $A$ to a point $B$, we essentially have to pass through all
points in between and the behavior of this network is much that of a
lattice with $\ell\sim N^\alpha$, although the diameter is here
smaller probably due to the existence of some rare longer links (in
the case of a lattice $\alpha=1/d$).


{\it Total cost and optimization}. Finally, we compute the average per node 
of the distances of all links
\begin{equation}
S=\frac{1}{N}\sum_{links} d_{ij}
\end{equation}
This quantity is a simple measure of the total cost of the
network. For $\eta\ll1$, preferential attachment is irrelevant and the
links distance is essentially distributed according to
$p(r)\sim\exp(-r/r_c)$ which leads to $S\sim L\eta$. In the opposite
situation $\eta\gg 1$, only preferential attachment is important and
the links distance is distributed according to $p(r)\sim
d\pi^{d/2}r^{d-1}/\Gamma(d/2+1)$ which gives for $d=2$ $S\simeq
L\pi/6$ (figure~\ref{ell-cost}b). When $\eta$ is increasing the total
cost is thus increasing while $N^*$ is decreasing. This implies that
for fixed $N$, the network which simultaneously minimizes the total
cost and the average diameter (ie. with small $N^*$) is non-trivial
and lies in between the `pure' scale-free network (ie. without
distance effect) and the `pure' spatial network (ie. with no
preferential attachment).


In summary, our results demonstrate the general importance of a cost
in the formation of networks which induces a dependence of all
quantities on the node density. In addition, the optimal network which
minimizes both the total length and the diameter (at fixed number of
nodes) will lie in between the scale-free and the spatial
network. This result could thus provide a simple explanation to the
large clustering coefficient and the positive correlations between
node degrees as it is observed in some cases where creating a long
link is costly such as social or transportation networks for example.


Acknowledgments. I thank Carl Herrmann and Paolo Provero for useful
and stimulating discussions. I also thank for its hospitality the
physics department of Torino-INFN where part of this work was
performed.





\end{multicols}


\begin{references}


\bibitem{Helmy02} A.~Helmy, online archive: cs.NI/0207069.

\bibitem{Nemeth02} G.~Nemeth and G.~Vattay, cond-mat/0211325.

\bibitem{Gorman03} S.P.~Gorman and R. Kulkarni, submitted to {\it
Environment and Planning Journal B} (2003).

\bibitem{Lakhina02} A.~Lakhina, J.B.~Byers, M.~Crovella, and I.~Matta, technical report, 
online version available at :
http://www.cs.bu.edu/techreports/pdf/2002-015-internet-geography.pdf

\bibitem{Simon55} H.A.~Simon, Biometrika {\bf 42}, 425 (1955).

\bibitem{Barabasi99} A.-L.~Barab\'asi and R.~Albert, Science {\bf
286}, 509 (1999).

\bibitem{Yook02} S.-H.~Yook, H.~Jeong, and A.-L.~Barabasi, PNAS
{\bf 99}, 13382 (2002).



\bibitem{Rozen02} A.F.~Rozenfeld, R.~Cohen, D. ben-Avraham, and
S.~Havlin, Phys. Rev. Lett. {\bf 89}, 218701 (2002).

\bibitem{Warren02} C.P.~Warren, L.M.~Sander and I.M.~Sokolov,
Phys. Rev. E {\bf 66}, 056105 (2002).


\bibitem{Dall02} J.~Dall and M.~Christensen, Phys. Rev. E {\bf 66},
016121 (2002).

\bibitem{Sen02} P.~Sen, K.~Banerjee, and T.~Biswas, Phys. Rev. E {\bf
66}, 037102 (2002).

\bibitem{Jost02} J.~Jost and M.P.~Joy, Phys. Rev. E {\bf 66}, 036126
(2002).


\bibitem{Manna02} S.S.~Manna and P.~Sen, cond-mat/0203216.

\bibitem{Xulvi02} R.~Xulvi-Brunet and I.M.~Sokolov, Phys. Rev. E {\bf
66}, 026118 (2002).


\bibitem{Clark51} C.~Clark, J.~R.~Stat.~Soc.~Ser. A {\bf 114}, 490 (1951). 
H.A.~Makse et al, Phys. Rev. E {\bf 58}, 7054 (1998).


\bibitem{Newman02a} M.E.J.~Newman, cond-mat/0205405.

\bibitem{Klemm02} K.~Klemm and V.M.~Eguiluz, Phys. Rev. E. {\bf 65},
057102 (2002).

\bibitem{Waxman88} B.M.~Waxman, IEEE {\it J. Select. Areas. Commun.}
{\bf 6}, 1617 (1988).

\bibitem{Berg02} J.~Berg and M.~Lassig, Phys. Rev. Lett. {\bf
89},228701 (2002).

\bibitem{Watts98} D.~Watts and S.H.~Strogatz, Nature {\bf 393}, 440
(1998).

\bibitem{Note_Poisson} In the case of a uniform density, the average
number of nearest neighbors is equal to $6$ for $d=2$. See for
example, C.~Itzykson and J.-M.~Drouffe, {\it Statistical Field
Theory}, Vol. 2 (Cambridge University Press, 1988).

\bibitem{Bollobas} B.~Bollobas, {\it Random Graph} (Academic Press,
New York 1985).

\bibitem{Callaway01} D.S.~Callaway, J.E.~Hopcroft, J.M.~Kleinberg,
M.E.J.~Newman, and S.H.~Strogatz, Phys. Rev. E {\bf 64}, 041902
(2001).

\bibitem{Newman02b} M.E.J.~Newman, cond-mat/0209450.

\bibitem{Vasquez02} A.~Vasquez and Y.~Moreno, to appear 
in Phys. Rev. E, Cond-mat/0209182.

\end{references}
\end{document}